\documentclass[aps,prb,twocolumn,superscriptaddress,longbibliography]{revtex4-1}
\usepackage[colorlinks=true,citecolor=blue,linkcolor=blue,breaklinks=true]{hyperref}

\usepackage{color,graphicx}
\usepackage{bm}
\usepackage{amsmath}
\usepackage{amssymb}
\usepackage{times}

\begin{document}

\newcommand {\ba} {\ensuremath{b^\dagger}}
\newcommand {\Ma} {\ensuremath{M^\dagger}}
\newcommand {\psia} {\ensuremath{\psi^\dagger}}
\newcommand {\psita} {\ensuremath{\tilde{\psi}^\dagger}}
\newcommand{\lp} {\ensuremath{{\lambda '}}}
\newcommand{\A} {\ensuremath{{\bf A}}}
\newcommand{\Q} {\ensuremath{{\bf Q}}}
\newcommand{\kk} {\ensuremath{{\bf k}}}
\newcommand{\qq} {\ensuremath{{\bf q}}}
\newcommand{\kp} {\ensuremath{{\bf k'}}}
\newcommand{\rr} {\ensuremath{{\bf r}}}
\newcommand{\rp} {\ensuremath{{\bf r'}}}
\newcommand {\ep} {\ensuremath{\epsilon}}
\newcommand{\nbr} {\ensuremath{\langle ij \rangle}}
\newcommand {\no} {\nonumber}
\newcommand{\up} {\ensuremath{\uparrow}}
\newcommand{\dn} {\ensuremath{\downarrow}}
\newcommand{\rcol} {\textcolor{red}}
\newcommand{\gcol} {\textcolor{green}}
\newcommand{\bcol} {\textcolor{blue}}
\newcommand {\beq} {\begin{equation}}
\newcommand {\eeq} {\end{equation}}
\newcommand {\bqa} {\begin{eqnarray}}
\newcommand {\eqa} {\end{eqnarray}}

\begin{abstract}

We study the transition from a many-body localized phase to an ergodic
phase in spin chain with correlated random magnetic fields. Using multiple
statistical measures like gap statistics and extremal entanglement
spectrum distributions, we find the phase diagram in the
disorder-correlation plane, where the transition happens at
progressively larger values of the correlation with increasing values
of disorder. We then show that one can use the average of sample variance of
magnetic fields as a single parameter which encodes the effects of the
correlated disorder. The distributions and averages of various
statistics collapse into a single curve as a function of this
parameter. This also allows us to analytically calculate the phase
diagram in the disorder-correlation plane.
 \end{abstract}
 \title{Many-body localized to ergodic transitions in a system with correlated disorder}
 \author{Abhisek Samanta}\email{abhiseks@campus.technion.ac.il}
\affiliation{ Physics Department, Technion, Haifa 32000, Israel}
\author{Ahana Chakraborty}\email{ahana@physics.rutgers.edu}
  \affiliation{Department of Physics and Astronomy, Center for Materials Theory,
Rutgers University, Piscataway, NJ 08854, USA}
\affiliation{Max Planck Institute for the Physics of Complex Systems, N\"othnitzer Str. 38, 01187, Dresden, Germany}
\author{Rajdeep Sensarma}\email{sensarma@theory.tifr.res.in}
 \affiliation{Department of Theoretical Physics, Tata Institute of Fundamental
 Research, Mumbai 400005, India}

\date{\today}

\maketitle

{\bf Introduction:} Strongly disordered interacting systems in one
dimension exhibit the phenomenon of many-body localization (MBL),
characterized by absence of transport and breakdown of standard
equilibrium statistical mechanics~\cite{mirlin,basko}. This has now been established both
theoretically~\cite{huse_review,altman_review,fabien_review} and through experiments on ultra-cold atomic systems~\cite{mblexpt1,mblexpt2}. The
energy eigenstates of these systems (at
finite energy density) exhibit low entanglement entropy, violate the
eigenstate thermalization hypothesis (ETH)~\cite{deutsch,srednicki,altman_review,huse_review,fabien_review}, and show characteristic
bimodal distribution of entanglement eigenvalues~\cite{yang,ent_spectrum}. They also show
characteristic features in the distribution of the lowest entanglement
spectra~\cite{abhisek_kedar}. These features clearly distinguish the MBL phase from the
standard ergodic phase, which follow ETH, have large entanglement
entropy, and show unimodal distribution of entanglement
eigenvalues. By increasing the fluctuations of the random couplings, which
are the microscopic manifestation of disorder, one can tune a system
from the ergodic phase to the MBL phase. This is known as the
MBL-ergodic phase transition ~\cite{oganesyan,apal,heisenberg,xxz,pollmann,nicolas,nicolas,dimer2d,quasiperiodic}.

The theoretical models of MBL~\cite{huse_review,altman_review,fabien_review} have mostly used local Hamiltonians with
random short-range couplings. These couplings are drawn independently from a
distribution, leading to models with uncorrelated disorder. However, in a
real experimental system, one would expect the random
couplings to be correlated; e.g. in a solid state system with
impurities, the disorder potential will decay with distance,
leading to correlated disorder. For non-interacting systems,
correlation in disorder is known to change the phenomenology of Anderson
localization~\cite{izrailev,croy}, with mobility edges appearing in systems that are
completely localized in absence of correlations.

In this Letter, we
explore the MBL-ergodic transition in a disordered spin chain in
presence of correlated disorder in magnetic fields. We first construct a model of
correlated disorder, where one can tune the overall scale of
fluctuations and correlations in the disorder independently. This
allows us to consider different scenarios ranging from weak to strong disorder with
negligible to large correlations and
tune between these limits using only two parameters. We track the
MBL-ergodic transition in the fluctuation-correlation plane using
several statistics from average value of the ratio of successive level
spacings of the eigenstates to the distribution of the lowest few
entanglement spectra of the eigenstates and find the corresponding
phase diagram. The ergodic phase exists at all values of correlation
at low disorder, while it exists only at very high correlations at
strong disorder.

\begin{figure*}
\centering
\includegraphics[width=0.45\textwidth]{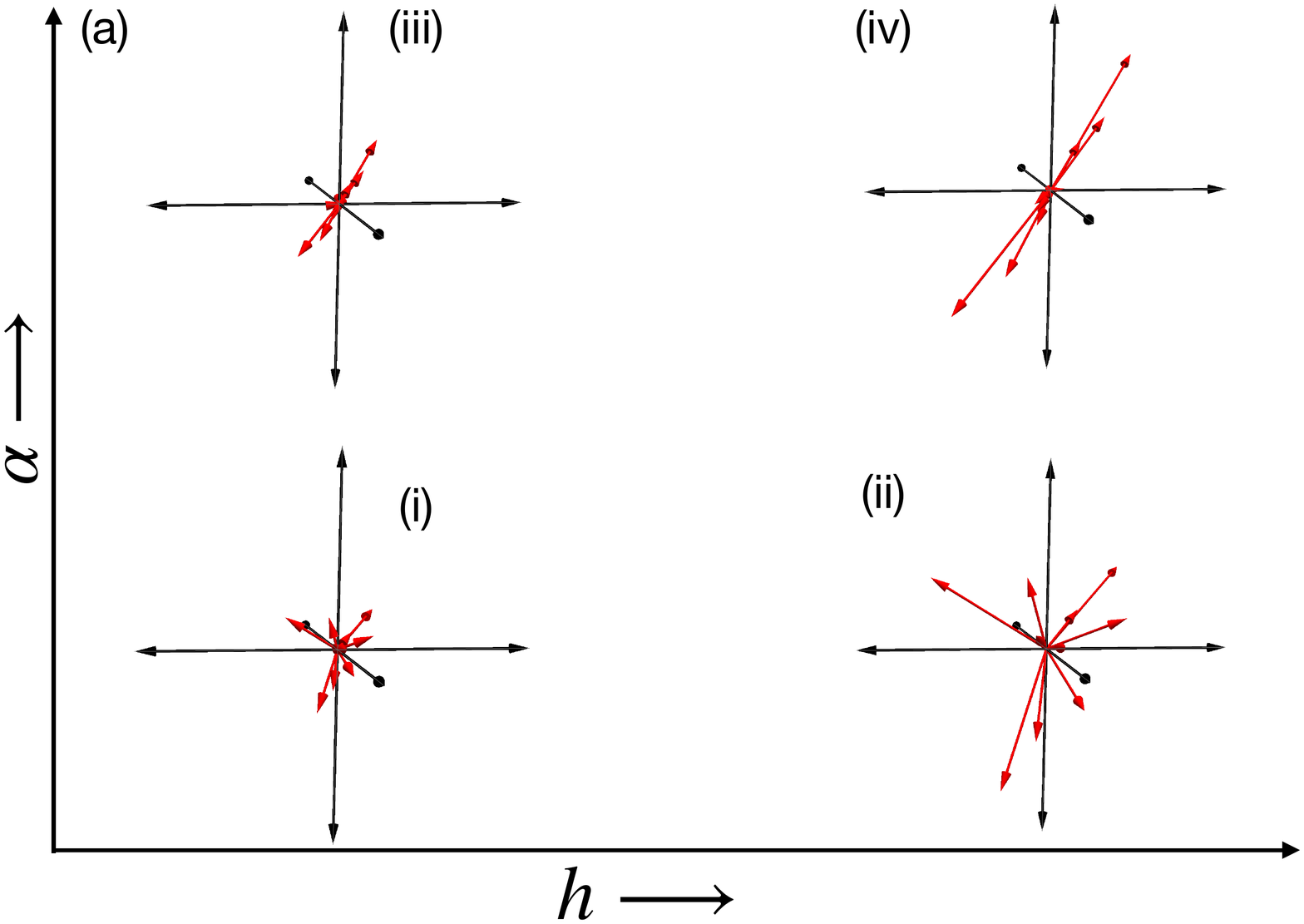}~~~~
\includegraphics[width=0.49\textwidth]{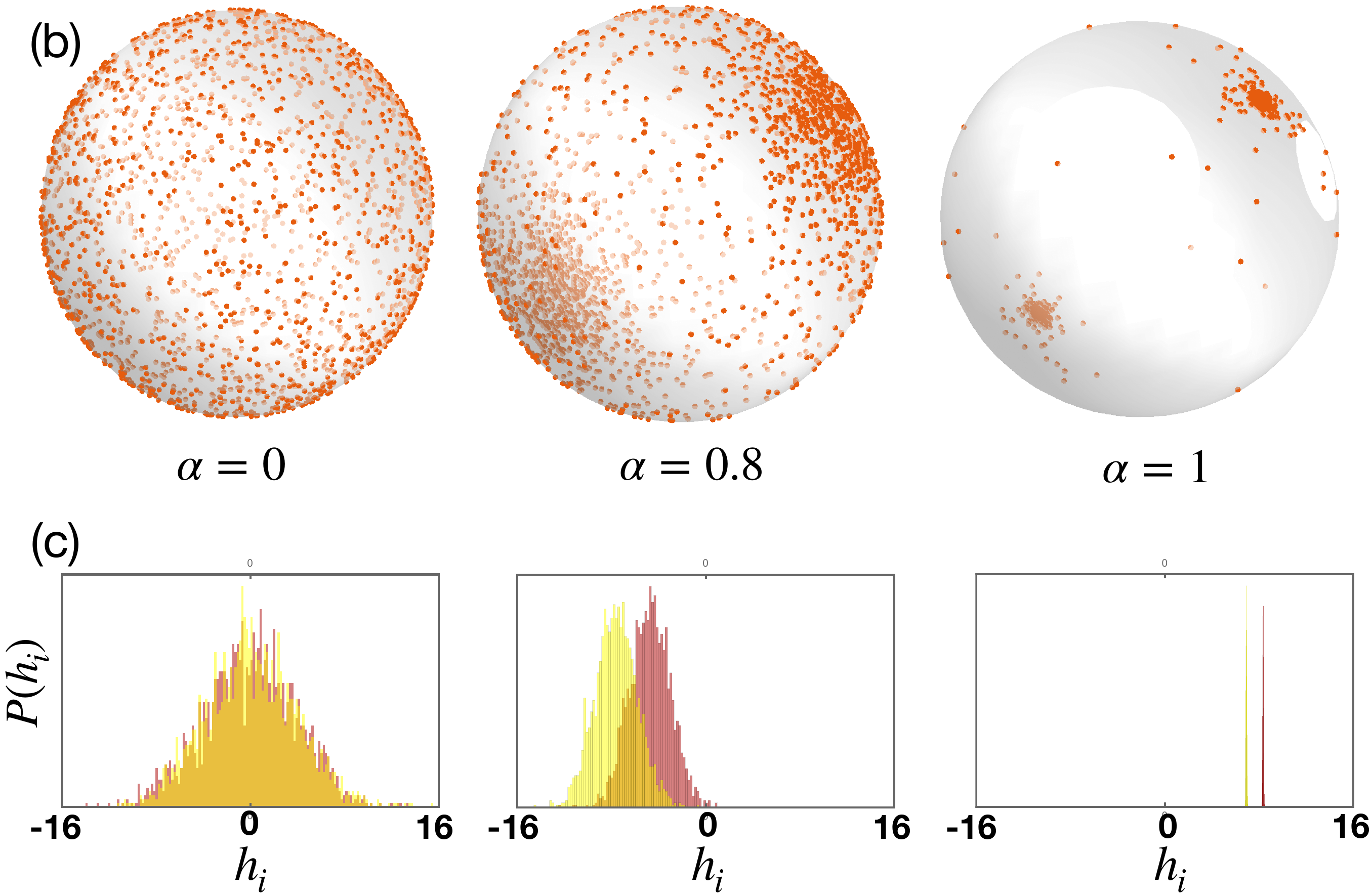}
\caption{ (a):  A schematic showing how the distribution of the
  random coupling vectors change with $h$ and $\alpha$. Few sample vectors 
  are shown for a $L=3$ site system in different regions of the $h-\alpha$
  plane: (i) $h=1.5$, $\alpha=0.1$, (ii) $h=3.5$, $\alpha=0.1$, (iii)
  $h=1.5$, $\alpha=0.99$ and (iv) $h=3.5$, $\alpha=0.99$. As $h$
  increases, the distribution of length of the vectors broadens,
  while increasing $\alpha$ narrows their angular distribution. (b): A
  representation of the unit vectors obtained from the random coupling
  vectors of the $L=3$ site system for three different values of
  $\alpha =0, ~ 0.8,~ 1$ at fixed $h=4$. This
  clearly shows the evolution of the angular distribution from an
  uniform distribution on the unit sphere at $\alpha=0$ to a narrowly
  bunched distribution around the $(1,1,1)$ direction. Note that the
  angular distribution has finite width at $\alpha=1$ since we are
  looking at $L=3$. This will approach a delta function as $L
  \rightarrow \infty$. (c) The distribution of magnetic fields $h_i$,
  obtained from a single sample realization, for a system of $L=2000$
  sites for $\alpha =0, ~ 0.8,~ 1$ at fixed $h=1$. The sample
  distribution narrows to a delta function as $\alpha$ is increased
  from $0$ to $1$, showing that fields at different sites in the
  system are getting correlated. Distributions obtained from two
  different samples are plotted in the same graph. At $\alpha=0$ the
  distributions overlap, while the overlap decreases with increasing
  $\alpha$. This shows that at large $\alpha$ the inter-sample
  variation of $h_i$ is much larger than its intra-sample variation.}
\label{Fig1disorder}
\end{figure*}

Increasing correlations in our model of disorder leads to less
fluctuations between random couplings in a given realization. Using
this idea, we define the average of sample variance as a single
parameter which controls the phases and phase boundaries of
the system. This is corroborated by the collapse of the gap statistics
and the lowest entanglement spectrum distributions, obtained at
different points in the fluctuation-correlation plane, into a single
curve when the data is plotted as a function of the average sample
variance. This allows us to analytically determine the phase boundary
of the system. We note that the correlated disorder we consider is
among the random couplings in real space, which is different from
considerations of correlated randomness in Fock space
~\cite{subroto,sthitadhi1,sthitadhi2}. In fact, while correlated
disorder in Fock space is thought to be essential for MBL phases to
exist~\cite{sthitadhi3}, increasing correlations in our disorder model
actually favours the ergodic phase.

{\bf The model:} We work with the spin-$1/2$ Heisenberg model on a one-dimensional chain of $L$ sites with nearest neighbour antiferromagnetic coupling. The spin-$1/2$ degrees of freedom also experience a random magnetic field along the $z$ direction at each site, which breaks both translation invariance and the global $SU(2)$ invariance of the underlying Heisenberg Hamiltonian. The Hamiltonian of this disordered model is given by
\beq
H = J \sum_i \sigma^z_i\sigma^z_{i+1}+\frac{1}{2} (\sigma^+_i\sigma^-_{i+1}+\sigma^-_i\sigma^+_{i+1}) -\sum_i h_i \sigma^z_i
\eeq
We set $J=1$ as a unit of energy. The magnetic fields $h_i$ are random variables drawn from a multivariate Gaussian distribution
\beq
P(h_1, h_2,..h_L) =\frac{1}{(2\pi)^{\frac{L}{2}}\sqrt{\textrm{Det} C^{-1}}}e^{-\frac{1}{2} h_i C^{-1}_{ij}h_j}
\eeq
 where the covariance matrix $C_{ij} = \langle h_ih_j\rangle$ is given by
 \beq
 C_{ij} =h^2\left [\delta_{ij} + \alpha (1-\delta_{ij})\right] ~~~ 0\leq \alpha \leq 1
 \eeq
 The parameter $h$ denotes the standard deviation of $h_i$, while their cross-correlators are given by $\langle h_jh_j\rangle =h^2\alpha$ for $i\neq j$. It is evident that $\alpha=0$ corresponds to the case of $h_i$ being uncorrelated Gaussian random variables, which was studied in Ref.~\onlinecite{abhisek_kedar}. In this case, the system exhibits a MBL to ergodic phase transition at a critical disorder strength $h_c(0)=1.6-1.8$, depending on the statistics used to track the transition.

\begin{figure}
	\centering
	\includegraphics[width=0.49\textwidth]{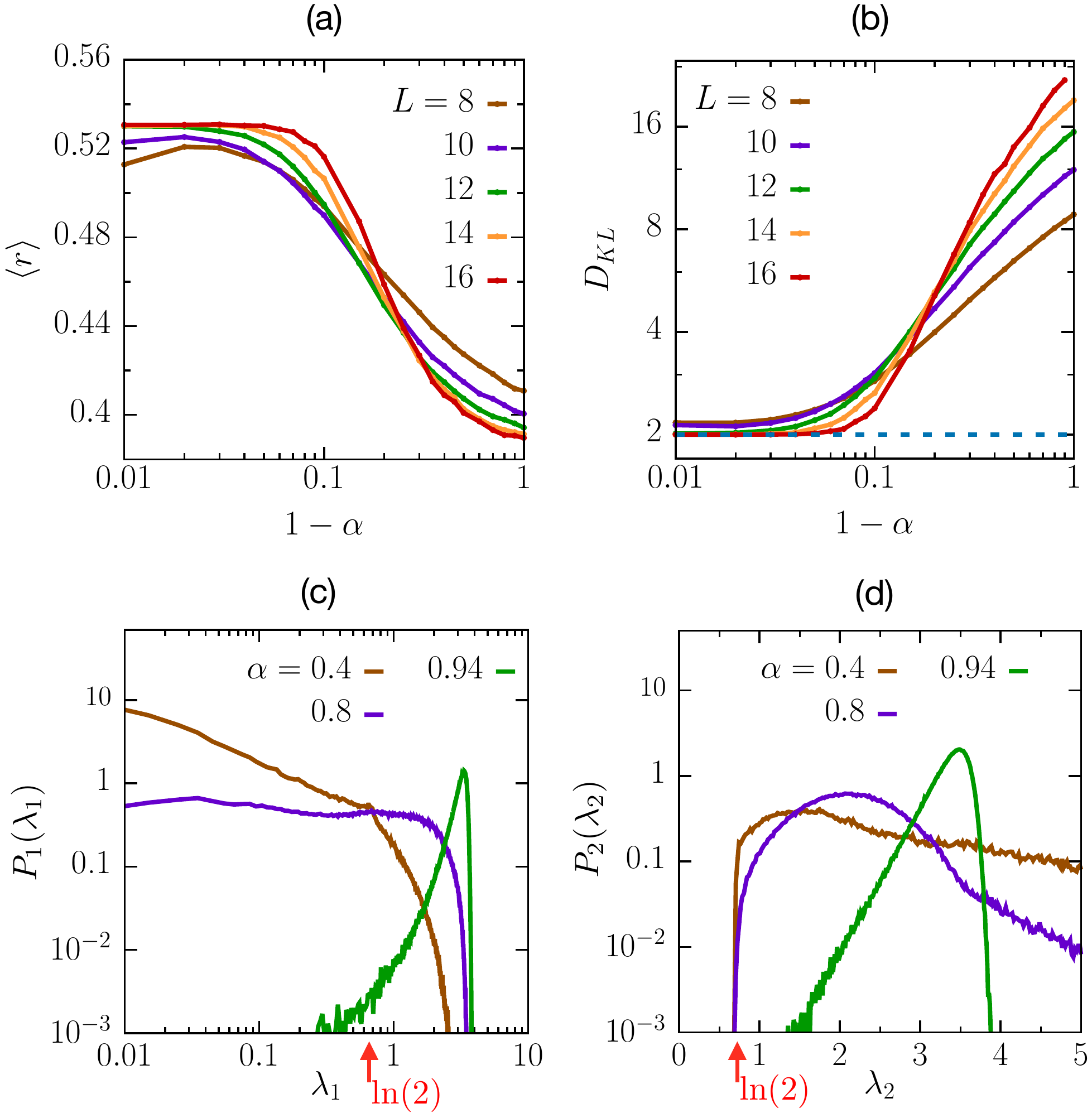}
	\caption{ (a)-(b): For the Heisenberg model with random
            magnetic fields drawn from a Gaussian distribution with correlation $\alpha$, 
           the average ratio of adjacent many-body gaps $\langle
            r\rangle$ (a) and the Kullback-Leibler parameter $D_{KL}$
            (b) are plotted as function of $1-\alpha$ for different
            system sizes. As $\alpha$ changes from $0$ (uncorrelated
            disorder) to $1$ (extremely correlated disorder), the
            system transitions from the MBL to the ergodic phase. The
            value of $\langle r\rangle$ changes from its Poissonian
            value of $0.39$ to the Wigner-Dyson value of
            $0.53$. $D_{KL}$ goes to a constant value of $2$ in the
            ergodic limit ($\alpha=1$), and scales with system size in
            the MBL limit ($\alpha\rightarrow 0$). (c)-(d): The
            distribution of lowest two entanglement eigenvalues,
            $\lambda_1$ and $\lambda_2$ are plotted for $\alpha=0.4,
            0.8$ and $0.94$. Here the system size $L=16$ and the
            subsystem size $L_A=8$. In the MBL phase ($\alpha=0.4$),
            $P_1(\lambda_1)$ shows a decreasing power law up to $\ln(2)$ and
            $P_2(\lambda_2)$ develops a finite value at $\ln(2)$. In
            the ergodic phase ($\alpha=0.94$),  $P_1(\lambda_1)$ goes
            to $0$ at $\lambda_1=0$, while  $P_2(\lambda_2)$ goes to
            $0$ at $\lambda_2=\ln 2$. $\alpha=0.8$ shows the
            distributions close to the MBL-ergodic transition. The
            standard deviation of the disorder has been kept fixed at
            $h=4$. } 
	\label{Fig2alphatransition}
\end{figure}
\begin{figure*}
	\centering
	\includegraphics[width=0.8\textwidth]{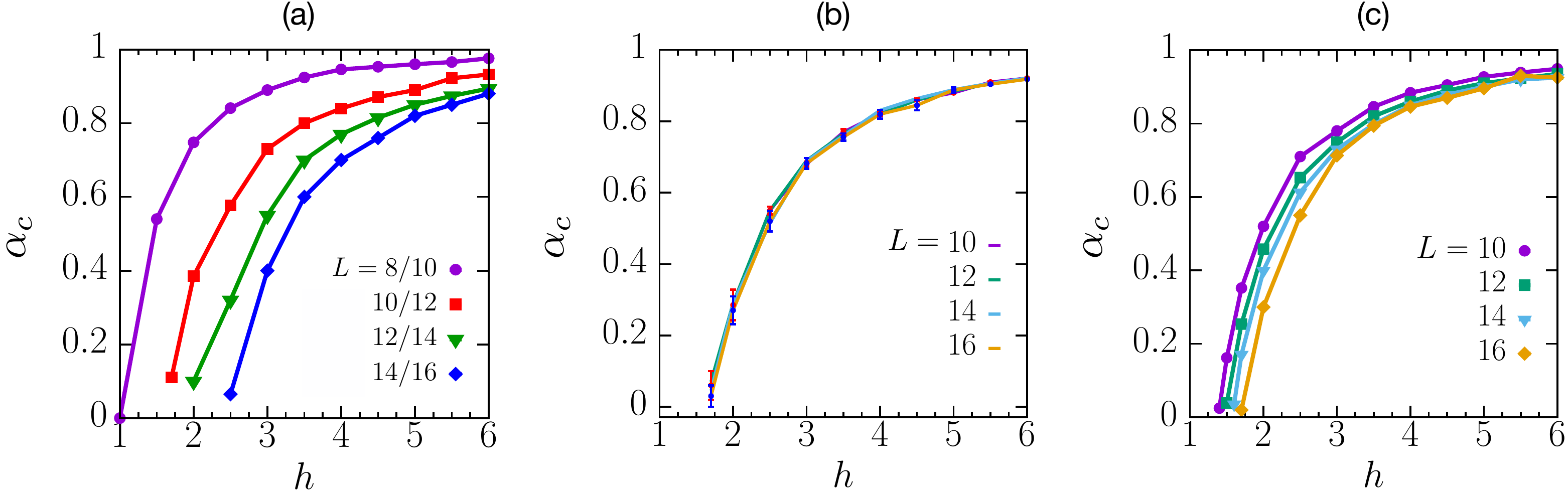}
	\caption{ Phase diagram in $h-\alpha$ plane, where the
          critical correlation $\alpha_c(h)$ for the transition has
          been obtained for each value of $h$ from (a) the crossing
          point between $\langle r\rangle$ vs $\alpha$ curves of
          successive system sizes, (b) the location of the sign change
          in the power law exponent $b$ in $P_1(\lambda_1)$ for
          different system sizes, and (c) the location where
          $P_2[\lambda_2=\ln(2)]$ vanishes for different system
          sizes. We notice strong system size dependence in (a), while
          the weakest system size dependence in using the criterion
          (b). In (b), the errorbars are obtained from errors in
          estimation of the exponent $b$.}
	\label{Fig3Phasediag}
\end{figure*}
 At finite $\alpha$, the $h_i$s are correlated random variables. To
 see how the distribution changes with $\alpha$, it is instructive to
 look at the eigenvectors and eigenvalues of the covariance
 matrix. The eigenvectors represent the linear combinations of $h_i$s
 which are uncorrelated Gaussian random variables with the variances
 given by the eigenvalues. For our choice, $(1,1, ....1)$ is an
 eigenvector with eigenvalue $h^2[1+(L-1)\alpha]$ and the rest of the
 degenerate eigenvalues are equal to $h^2(1-\alpha)$ with the
 eigenvectors lying in a $L-1$ dimensional subspace orthogonal to
 $(1,1,...1)$. As $\alpha \rightarrow 1$, the probability
 distributions along these orthogonal directions shrink to a delta
 function , i.e. all the coupling vectors lie along the $(1,1,...1)$
 direction. This is the extremely correlated limit, where the magnetic
 fields in each realization are exactly same on all lattice sites. The
 value of this common magnetic field varies from one realization to
 another.

A schematic representation of how the distribution of the vector of
couplings change with $h$ and $\alpha$ is shown in
Fig.~\ref{Fig1disorder} (a) in the $h-\alpha$ plane (for $L=3$). Increasing $h$ at
fixed $\alpha$ broadens the distribution of the length of these vectors while keeping
their angular distribution same. Similarly increasing $\alpha$ while
keeping $h$ fixed causes the angular distribution to narrow around the
$(1,1,1)$ direction. In Fig.~\ref{Fig1disorder} (b) we plot the unit
vectors corresponding to different sample realizations of the coupling
vector on a unit sphere for $\alpha=0,~0.8, 1.0$. The change of the angular distribution from
an uniform distribution at $\alpha=0$ to a narrow distribution as
$\alpha \rightarrow 1$ is clearly seen here. In Fig~\ref{Fig1disorder}
(c), we plot the distribution of individual $h_i$s obtained from a
single sample for a system of size $L=2000$ for the same
values of $\alpha$ as in Fig.~\ref{Fig1disorder} (b). We plot two
different distributions corresponding to two different samples. When
the disorder is uncorrelated ($\alpha=0$), the distributions overlap,
showing that the samples are statistically similar to each other. As
$\alpha$ is increased, the overlap decreases, showing that couplings
in a sample are more narrowly distributed than couplings taken from
different samples. In the extreme limit of $\alpha=1$, we find that
all couplings in a sample are same, but their value varies from one
sample to another.

Our correlated disorder model has the simplicity that it can be tuned from uncorrelated to extremely correlated limit using a single parameter. Other models, where the covariance $C_{ij}$ depends on the separation between $i$ and $j$, introduce additional lengthscales in the problem, and tuning them from uncorrelated to extremely correlated limits require going to large system sizes, which are numerically not accessible. 

{\bf MBL-ergodic transition with varying $\alpha$:} We will first consider
the fate of this system as $\alpha$ is tuned from $0$ to $1$ at a
fixed value of $h$. We will focus our attention on the system
at $h=4$, where the uncorrelated system ($\alpha=0$) is in a many-body localized phase. As we increase $\alpha$, the system undergoes a
phase transition from the MBL to ergodic phase. We numerically
diagonalize the Hamiltonian for a large number of disorder
realizations and use the eigenvalues $E_n$ and eigenstate wavefunctions $\phi_n(i)$
in the middle third of the spectrum to construct several statistics to look at this transition:
\begin{figure}[h!]
	\centering
	\includegraphics[width=0.485\textwidth]{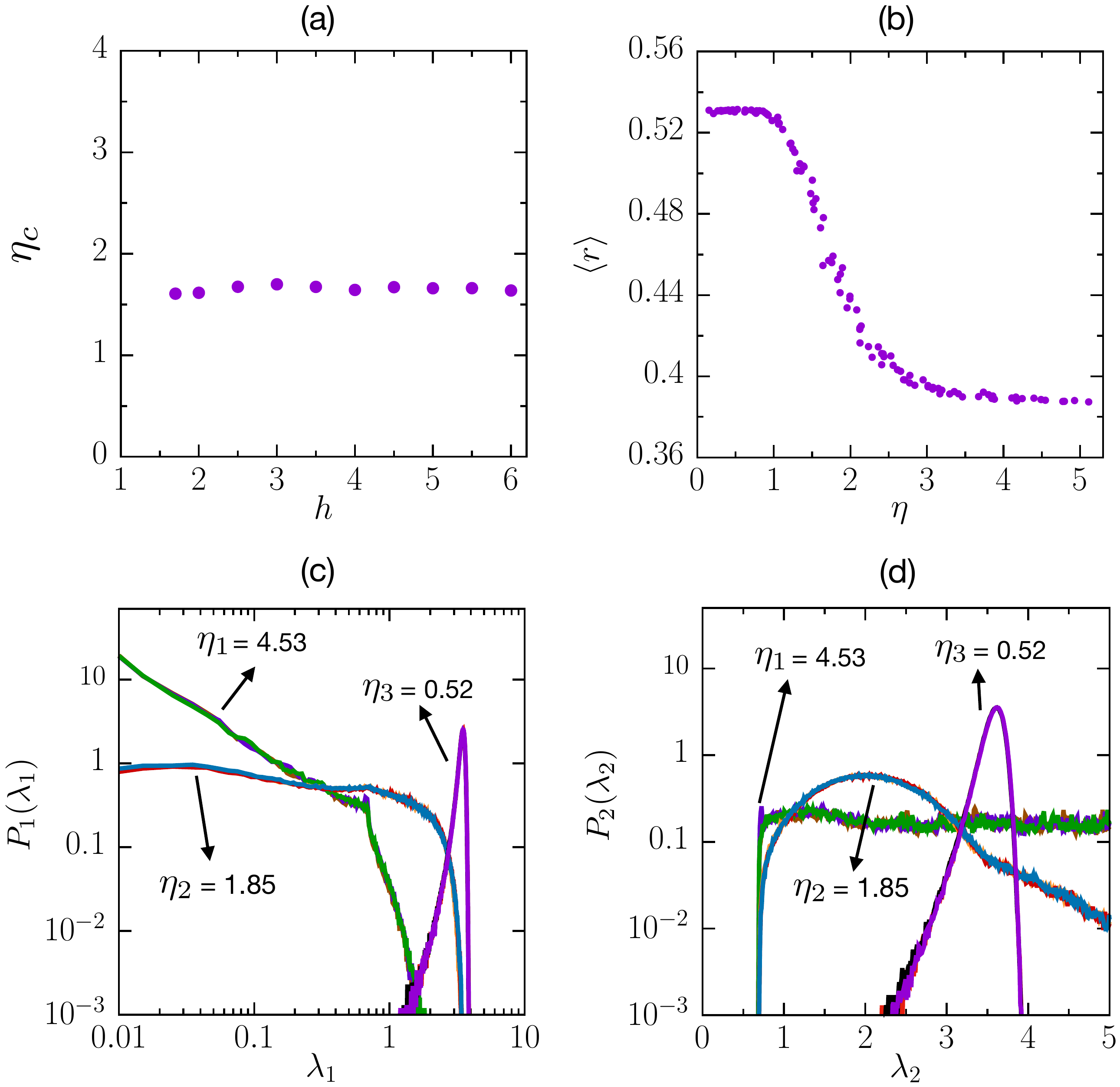}
	\caption{ (a) The MBL-ergodic phase boundary in the
          $h-\alpha$ plane can be understood using a single
          parameter, the average sample variance $\eta^2$. The value of
          $\eta$ at the transition point, $\eta_c$, is plotted as a
          function of $h$. The constant $\eta_c(h)$ indicates that the
          MBL-ergodic transition is controlled by $\eta$. (b) $\langle
          r\rangle$ is plotted as a function of $\eta$ for several
          values of $(h,\alpha)$. The data collapses to a single
          curve (which changes from Wigner-Dyson to Poissonian value)
          with a small spread. (c)-(d): Distribution of (c) lowest
          entanglement eigenvalue $P_1(\lambda_1)$ and (d) second
          lowest entanglement eigenvalue $P_2(\lambda_2)$ are plotted
          for three values of $\eta$:  $\eta_1=4.53$ (MBL phase),
          $\eta_2=1.85$ (transition
          point) and $\eta_3=0.52$ (ergodic phase). The curve for each value of $\eta$ is composed of
          data from three sets of $(h,\alpha)$ values. All curves with the same $\eta$ collapse to a single curve.}
	\label{Fig4Oneparam}
\end{figure}

(i) {\it The gap ratio
   $\langle r\rangle $:}
 The level repulsion between the eigenstates can be parametrized by
 the ratio of the gap between successive energy eigenvalues, $r_n=(E_{n+1}-E_{n})/(E_n-E_{n-1})$. The gapmoment $\langle r \rangle$ is the average of this quantity, both over eigenstates in a
 disorder realization and over disorder realizations. $\langle r\rangle$ is expected
 to have an average value of $0.53$ in the
 ergodic phase corresponding to Wigner-Dyson distribution for level
 statistics, and an average value of $0.39$ in the MBL phase
 corresponding to a Poisson distribution for level statistics. In
 Fig.~\ref{Fig2alphatransition}(a), we plot $\langle r \rangle$ as a
 function of $(1-\alpha)$ for different system sizes from $L=8$ to $L=16$ and see that the
 system clearly shows a transition from MBL to ergodic phase as $\alpha$ is
 increased. The location of the transition point, as measured from
 crossing of curves for increasing system sizes, vary over a broad
 range, similar to the transition for uncorrelated disorder as a
 function of $h$~\cite{oganesyan,apal,abhisek_kedar}.

 (ii) {\it Kullback-Leibler divergences:} The projection of an
 eigenstate $|\psi_n\rangle$ on a set of complete basis states (say the eigenstates of
 $\sigma^z_i$, $|\mu\rangle =\otimes_i|\sigma^z_i\rangle$), defines a
 probability distribution $p^{(n)}_\mu=|\langle \psi_n|\mu\rangle|^2$. One can construct
 the Kullback-Leibler divergences of the probability distributions
 obtained from successive energy eigenstates, $D_{KL}=\sum_\mu
 p^{(n)}_\mu\left( \ln{p^{(n)}_\mu\over {p^{(n-1)}_\mu}}\right)$ to compare how
 similar the successive eigenstates are. $D_{KL}$,
 averaged over adjacent pairs of eigenstates, and over disorder
 realizations, is expected to scale with system size in the MBL phase
 and should go to a value of $2$ in the ergodic phase ~\cite{heisenberg}.
In Fig.~\ref{Fig2alphatransition}(b) we have plotted $D_{KL}$ as a
function of $(1-\alpha)$ for different system sizes. We notice that
$D_{KL}$ reaches a constant value 2 when $\alpha\rightarrow 1$
(ergodic limit), and it scales with the system size as $\alpha$
decreases (MBL phase)~\cite{heisenberg}.
Note that although $D_{KL}$ distinguishes the MBL phase from the
ergodic phase,
it is not a good indicator of the MBL-ergodic transition, since the
crossover from $D_{KL} \sim L$ to $D_{KL}=2$ takes place over a wide
range of $\alpha$ and is sensitive to the system size..

While the statistics described above show a gradual change from MBL to
ergodic behaviour as a function of $\alpha$, sharper distinctions
can be drawn from the distribution of entanglement spectra of reduced
density matrices constructed from the energy eigenstates of the
system~\cite{nandkishore,ent_spectrum,abhisek_kedar}. To see this, we consider 
a subsystem of size $L_A$ ($L_A=L/2$
for numerical data presented here). We start with the eigenstate $|\phi_n\rangle$,
trace out the degrees of freedom lying outside the subsystem to
construct the reduced density matrix $\hat{\rho}_A$. Here we look at the
distribution of the lowest two entanglement eigenvalues of $\ln\hat{\rho}_A$,
$\lambda_{1(2)}$:

(i) {\it $P_1(\lambda_1)$:} The lowest entanglement eigenvalue $\lambda_1$ has a power law
distribution in the range $0$ to $\ln(2)$, $P_1(\lambda_1) \sim
\lambda_1^{-b} $, with the exponent going from positive values in the
MBL phase to negative values in the ergodic phase. The distribution
shows characteristic kink at $\ln(2)$ in the MBL phase with an
exponential decay at large values, while the ergodic phase shows a
prominent peak at $\lambda_1 \sim 3.5$. In
Fig..~\ref{Fig2alphatransition}(c), we plot $P_1(\lambda_1)$ for three
values of $\alpha=0.4,0.8$, and 0.94 with a fixed $h=4$. We find that $\alpha
=0.4$ is in the deep MBL phase, while $\alpha=0.94$ is in the ergodic
phase. $\alpha=0.8$ shows the distribution at a point close to the
transition.

(ii)  {\it $P_2(\lambda_2)$:} The second lowest entanglement eigenvalue
$\lambda_2$ is always greater than $\ln(2)$. In the MBL phase the
distribution is finite at $\lambda_2= \ln(2)$, while the ergodic phase
shows exponentially small weight at this point. We note that a finite value of $P_2[\lambda_2=\ln(2)]$ corresponds to the exchange of exactly one bit information between the subsystem and its surroundings~\cite{abhisek_kedar}. In
Fig.~\ref{Fig2alphatransition}(d), we plot $P_2(\lambda_2)$ for three
values of $\alpha= 0.4,0.8,$ and 0.94 with a fixed $h=4$. From the
$P_2(\lambda_2)$ curves, we once again find that $\alpha
=0.4$ and $0.94$ corresponds respectively to systems deep in the MBL and ergodic phases, while the intermediate value $\alpha=0.8$ shows the distribution close to the
transition point.
\begin{figure*}[t!]
\centering
\includegraphics[width=0.95\textwidth]{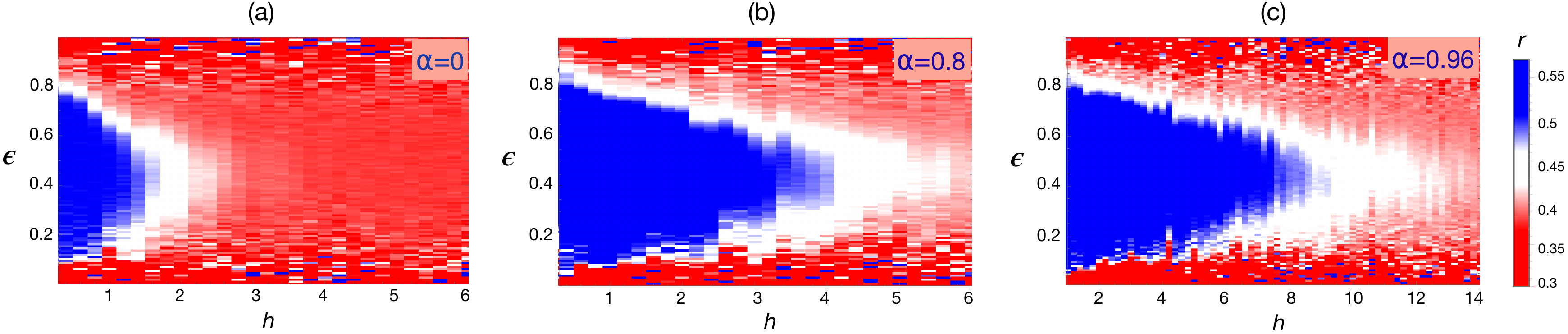}
\caption{The density-plots of the average ratio of adjacent gaps $r$
  are shown in energy-disorder ($\epsilon-h$) plane for three values
  of correlation (a) $\alpha=0$, (b) $\alpha=0.8$, and
  (c) $\alpha=0.96$. The energy $\epsilon$ is scaled by its range to lie
  between $0$ and $1$. With increase in $\alpha$, the blue region
  showing the ergodic states extend to larger values of $h$ and the
  transition from Wignar-Dyson to Poissonian statistics is broadened.}
\label{Fig5mobedge}
\end{figure*}
Additionally, we have also looked at the distribution of all
entanglement eigenvalues which changes from a unimodal distribution in
the ergodic phase to a bimodal distribution in the MBL phase and find
the same MBL to ergodic transition as a function of $\alpha$ (see Appendix A for details).

{\bf The phase diagram:} Having established that there is a
MBL-ergodic transition as a function of $\alpha$ at large enough $h$, we now focus on the phase diagram of the
system in the $h-\alpha$ plane, determined by fixing $h$ on a grid of
values and tuning $\alpha$ in each case. We use three different criterion, with
varying degree of system size dependence, to obtain the transition
point $\alpha_c(h)$ : (i) The crossing point between the $\langle r\rangle$ vs $\alpha$ curves of
successive system sizes. The phase diagram obtained from this
criterion is plotted in Fig.~\ref{Fig3Phasediag}(a) for different
system sizes. (ii) The value
of $\alpha$ where the exponent $b$ of the lowest entanglement
eigenvalue distribution $P_1(\lambda_1)$ changes sign. The phase diagram from this
criterion is plotted in Fig.~\ref{Fig3Phasediag}(b) for different
system sizes. (iii) The value
of $\alpha$ where $P_2[\lambda_2=\ln(2)]$ reaches $0$ (we use a threshold
value of $10^{-3}$). The corresponding phase diagram is plotted in
Fig.~\ref{Fig3Phasediag}(c). All three criteria give
qualitatively similar phase diagrams: at low $h < h_c(0)$, where
$h_c(0)$ is the critical disorder for the uncorrelated system, the system
is ergodic for all values of $\alpha$. For $h >h_c(0)$, there is a
$\alpha_c(h)$ below which the system is in MBL phase, while it is in
ergodic phase for $\alpha >\alpha_c(h)$. $\alpha_c(h)$ is a
monotonically increasing function of $h$, rising rapidly from $0$ just above
$h=h_c(0)$ and slowly saturating to $1$ as $h \rightarrow \infty$. In
the next section we will derive an analytic expression for
$\alpha_c(h)$. The phase boundaries obtained from the $\langle r\rangle$ vs $\alpha$
curves are strongly system size dependent, whereas the phase diagram
obtained from change of slope of $b$ has the weakest size
dependence. While the finite size systematics for the phase boundary
obtained through (i) and (iii) sets the leading error estimates,
we estimate the errors in phase boundary obtained by (ii) from errors
in estimating $b$ (see Appendix B for the estimation of error in $b$).

{\bf The effective one-parameter model:} If we consider the random
magnetic fields at different sites in a particular disorder
realization (or sample) as components of a
vector, the parameter $\alpha$ controls the variation between these
components. When $\alpha=1$, all the couplings in a particular sample are same; hence each
realization is translation invariant, and the system should behave
ergodically. The MBL vs ergodic behaviour of the system should thus be
controlled by a measure of variation of the couplings within a
sample. This can be captured by defining a
sample/realization variance $\sigma^2 = \frac{1}{L} \sum_ih_i^2
-\left( \frac{1}{L} \sum_ih_i\right)^2$. $\sigma^2$ is a random
variable, and its average over all disorder realizations is defined as
$\eta^2=\langle \sigma^2\rangle$. For uncorrelated disorder ($\alpha=0$), one
can easily show that for large system sizes, $\eta=h$ is just the
standard deviation of the fields. On the other hand, for $\alpha=1$,
$\sigma^2=0$ for every realization and hence $\eta=0$. For a generic
$\alpha$, one can show that (see Appendix C for
details)
\beq
\eta^2 =h^2(1-\alpha)
\label{etaalpha}
\eeq
We will now show that the phases and the phase diagram in
the $h-\alpha$ plane can be understood in terms of this single
parameter $\eta$. We first plot the value of $\eta$ at the phase
transition points, $\eta_c(h)=h\sqrt{1-\alpha_c(h)}$, as a function of $h$
in Fig.~\ref{Fig4Oneparam}(a). For $h>h_c(0)$, $\eta_c$ is constant, showing
that the phase diagram can be understood in terms of this single
parameter. Since $\eta_c=h_c(0)$ (for $\alpha=0$), one can use this to
analytically obtain the phase boundaries in the $h-\alpha$ plane,
\bqa
\no \alpha_c(h) &= 1-\left(\frac{h_c(0)}{h}\right)^2& ~~\text{for}~~  h \geq
h_c(0)\\
& =0 & ~~\text{for}~~  h <
h_c(0)
\eqa
Beyond the phase boundary, we now show that the behaviour of the
system in the two phases across the transition is a function of a
single parameter $\eta$. To see this, in Fig. ~\ref{Fig4Oneparam}(b), we plot
the gapmoment $\langle r\rangle$ obtained at different $(h,\alpha)$ points as a
function of $\eta$ and show that the data collapses to a single curve
with a small spread. This data collapse is even more remarkable if one
looks at the distribution of lowest entanglement eigenvalues. In Fig
~\ref{Fig4Oneparam}(c), we plot $P_1(\lambda_1)$ for three different values of
$\eta$, in the MBL, ergodic and the critical phase. Each curve is
actually three curves (three different $(h,\alpha)$ values
corresponding to the same value of $\eta$). These three curves are
indistinguishable due to data collapse. Similar trends are seen in
$P_2(\lambda_2)$ plotted in Fig ~\ref{Fig4Oneparam}(d). This conclusively shows
that there is a single parameter $\eta$ which is relevant for
understanding the phases and the phase transitions in the $h-\alpha$ plane.

{\bf Mobility edge:} Long-range correlations in disorder models are
known to lead to mobility edges, even in non-interacting
systems~\cite{ganeshan,modaksubroto1}. While the question of a many-body mobility
edge has been debated~\cite{sankar,heisenberg,modaksubroto2,
  monika,mondaini}, for spin chains with uncorrelated disorder, 
it has been shown~\cite{heisenberg} that at weak disorder, there is an energy range
over which the gap statistics changes from its Wigner-Dyson
distribution value of $0.51$ to its value in the Poisson limit
($0.39$). Due to finite size of the system, the change happens over a
finite energy window. At large disorder, all states are many-body
localized, leading to the disappearance of the mobility edge. In
Fig.~\ref{Fig5mobedge}(a)-(c), we plot the gap statistics $r$ as a color
plot in the energy-$h$ plane for three different values of
correlation, $\alpha=0, 0.8$ and $0.96$. In each case,
we see a transition at weak disorder, with the ergodic states at the
center of the spectrum extending
to larger values of $h$ as $\alpha$ grows. This simply reflects the
fact that the critical disorder increases with increasing $\alpha$. In
addition, the transition energy range becomes larger as correlation
$\alpha$ is increased, and a sharp ``edge'' is absent, as seen in Fig~\ref{Fig5mobedge}(a)-(c).

In summary, we have studied a spin chain with correlated disorder and have mapped
out the phase diagram in the disorder-correlation plane. We show that
the sample variance can be treated as a single parameter controlling
the phases and phase transitions in the system. Using this, we can
analytically find the phase boundary of the system. The usual picture
of ``mobility edges'' separating localized and delocalized states at
weak disorder in systems with uncorrelated disorder continues to hold in this case
with some small modifications.

 \medskip

\begin{acknowledgments}
The authors thank Subroto Mukerjee for useful discussions. The authors acknowledge computational facilities at Department of Theoretical Physics, TIFR Mumbai. A.S. also acknowledges 
the computational facilities of Physics Department, Technion. R.S. acknowledges support of the Department of Atomic Energy, Government of India, under Project Identification No. RTI 4002.
\end{acknowledgments}

\appendix
\section{Distribution of entanglement spectrum across the MBL-ergodic transition}
\label{App:Plambda}

\begin{figure}[h!]
	\centering
	\includegraphics[width=0.3\textwidth]{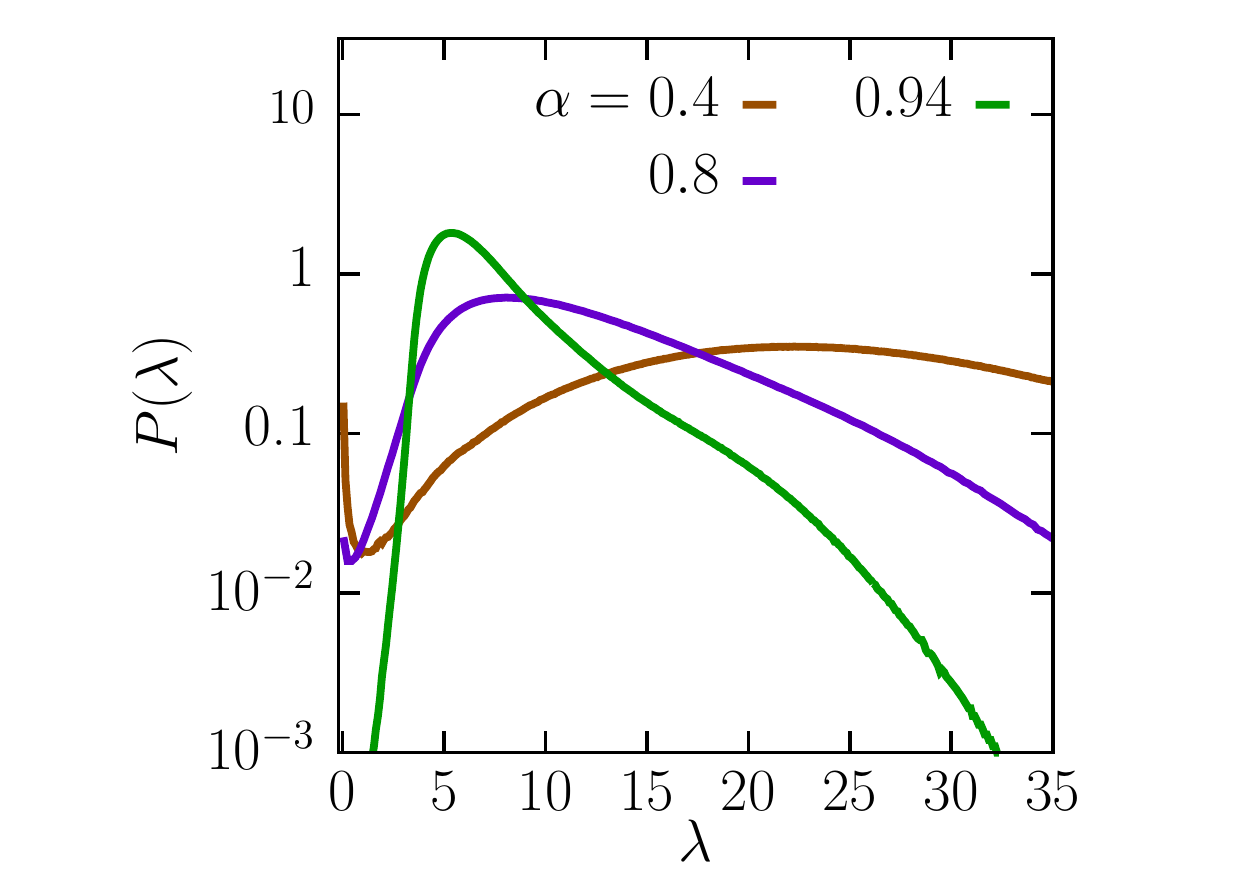}
	\caption{ The distribution of the full entanglement spectrum
          $P(\lambda)$ for system size $L=16$ and subsystem size
          $L_A=8$ for different values of $\alpha$ for fixed $h=4$. The distribution shows a peak in the ergodic phase ($\alpha=0.94$) near $\lambda\sim 8\ln(2)$, while it is characterized by two peaks in the MBL phase ($\alpha=0.4$). At $\alpha=0.8$ the system is near the transition. The appearance of the peak near $\lambda=0$ can be used to track the transition.}
	\label{SuppFig1}
\end{figure}

In the main text we have studied the distribution of the lowest two
  entanglement eigenvalues, $P_1(\lambda_1)$ and $P_2(\lambda_2)$ to
  distinguish MBL and ergodic phases. We have also shown that the
  power law exponent of $P_1(\lambda_1)$ below $\ln(2)$ and
  $P_2[\lambda_2=\ln(2)]$ can be used to track the transition
  accurately in the $\alpha-h$ plane. However one can also look at the
  whole entanglement spectrum $P(\lambda)$ to study the phase
  transition. In Fig.~\ref{SuppFig1} we show $P(\lambda)$ for three
  values of $\alpha$, i.e. $\alpha=0.4,0.8$ and $0.94$. The system
  size is $L=16$ and the subsystem size is $L_A=8$. In the ergodic
  phase all eigenvalues of the reduced density matrix scale as $\rho_n\sim
  1/2^{L_A}$, and hence we see a peak
  in the distribution for $\alpha=0.94$ around $\lambda\sim
  8\ln(2)\approx 5.5$. As the system moves from the ergodic to
  MBL phase, a large weight appears at $\lambda=0$ which corresponds
  to the occurrence of product states~\cite{abhisek_kedar}. Therefore,
  MBL phase is characterized by the existence of two peaks, one near
  $\lambda=0$ and another corresponding peak at large $\lambda$. The
  MBL-ergodic transition can be tracked by tracking the appearance of
  the $\lambda=0$ peak~\cite{nandkishore}. Using this metric, we see
  that we can reproduce the MBL-ergodic transition as a function of
  $\alpha$ discussed in the main text.

\section{Estimation of error in the phase diagram obtained from the power-law exponent $b$}
\label{App:Errorb}
To draw the phase diagram in the $h-\alpha$ plane, we have used three
different criteria in the main text. We have seen that the phase
boundary obtained from the sign change of the power-law exponent $b$
of $P_1(\lambda_1)$ has the weakest system size dependence. Therefore
it is useful to estimate the error in drawing the phase boundary using
this criterion. In Fig.~\ref{Fig3Phasediag}(b) of the main text we
have plotted the phase diagram and the corresponding error bars. Here
we provide the technical details in the estimation of errors. For a
given $h$, we move on a grid of $\alpha$ and fit $P_1(\lambda_1)$ for
each $\alpha$ in the range $0<\lambda_1<\ln(2)$ with a power law
$P_1(\lambda_1)\sim\lambda_1^{-b}$.  This gives us a standard error of
fitting in
$b$, $\Delta b$. We estimate $\alpha_c(h)$ by tracking the sign change
in $b$ as a function of $\alpha$. Similarly we estimate the error in
$\alpha_c(h)$ by tracking the sign change in $b+\Delta b$ and
$b-\Delta b$ as a function of $\alpha$. This is used to construct the
errorbars shown in Fig.~\ref{Fig2alphatransition}.
\section{Relation between $\eta, h$ and $\alpha$}
\label{App:eta}
In this Appendix we will calculate the relation between $\eta, h$ and $\alpha$ used in the main text. We consider a sample of $L$ random variables $h_i$ drawn from a correlated Gaussian distribution with standard deviation $h$ and correlation $\alpha$. We define $\bar h={1\over L}\sum_i h_i$  and $\sigma^2 = \frac{1}{L} \sum_ih_i^2
-\left( \frac{1}{L} \sum_ih_i\right)^2$ to be the mean value and the variance within the sample respectively. The average of sample variance over all disorder realization $\eta^2=\langle \sigma^2\rangle$ is given by,
\begin{align}
\eta^2 &= \frac{1}{L}\Big\langle\sum_{i=1}^L (h_i-\bar h)^2\Big\rangle \no\\
&= (h^2-\langle {\bar h}^2\rangle)
\label{etasqr}
\end{align}

In this case,
\begin{align}
\bar h^2 &={1\over L^2}\left[\sum_i^n h_i^2 +  h_1(h_2+h_3+...+h_L) \right. \\ 
&  \!\!\!\!\!\!+h_2(h_1+h_3+...+h_L)+...+h_L(h_1+h_2+...+h_{L-1})\Big], \no
\end{align}
Disorder average of $\bar h^2$ yields $\langle \bar h^2\rangle={1\over L^2}\left[L h^2+L(L-1)\alpha h^2\right]$. Using this in Eqn.~\ref{etasqr} 
we find $\eta^2=(1-\alpha)h^2+ {\cal O}(1/L)$, which reduces to
Eq.~\ref{etaalpha} in the main text for large $L$.

\bibliographystyle{unsrt}
\bibliography{refs_CorrelatedMBL.bib}
\newpage

\end{document}